\documentclass[acmsmall]{acmart}
\usepackage{etex}

\usepackage{subcaption} 

\acmConference[]{}

\usepackage{booktabs}   
\usepackage{subcaption} 
\usepackage[noend]{algpseudocode}

\usepackage{graphicx}
\usepackage{amsmath}
\usepackage{xspace}
\usepackage{footnote}
\usepackage{amsfonts}
\usepackage{natbib}
\usepackage{hhline}
\usepackage{tabularx}
\usepackage{scratch3}
\usepackage{diagbox}
\usepackage{multirow}
\usepackage{setspace}
\usepackage{epsfig}
\usepackage{url}

\usepackage{breakurl}
\usepackage{booktabs}
\usepackage{tabularx}
\newcolumntype{Y}{>{\raggedright\arraybackslash}X}
\usepackage{tikz}
\usetikzlibrary{arrows.meta,positioning,shapes.geometric,shapes.misc,fit,backgrounds}

\usepackage{xcolor}
\usepackage{lipsum}
\usepackage{courier}
\usepackage{listings}
\usepackage{caption}
\usepackage{colortbl}
\usepackage{arydshln} 
\usepackage{makecell} 
\usepackage{tikz}
\usepackage{color}
\usepackage{rotating}

\usepackage{varwidth}
\usetikzlibrary{fit,arrows,calc,positioning,quotes}

\usepackage{pgfplots}

\usepackage{mathtools}
\usepackage{fancyvrb}
\usepackage{textcomp}

\usepackage{longtable}
\usepackage{pdflscape} 
\usepackage{threeparttable} 
\usepackage{array}     

\usepackage{stmaryrd}

\usepackage{algorithm}
\usepackage[noend]{algpseudocode}

\usepackage{alltt}

\tikzset{
  -|-/.style={
    to path={
      (\tikztostart) -| ($(\tikztostart)!#1!(\tikztotarget)$) |- (\tikztotarget)
      \tikztonodes
    }
  },
  -|-/.default=0.5,
  |-|/.style={
    to path={
      (\tikztostart) |- ($(\tikztostart)!#1!(\tikztotarget)$) -| (\tikztotarget)
      \tikztonodes
    }
  },
  |-|/.default=0.5
}

\usetikzlibrary{shapes.geometric}

\long\def\com#1{}

\newcommand{\app}{\texttt{Raven}\xspace}

\newcommand{\para}[1]{\smallskip\noindent {\bf #1}}

\newcommand{\squishlist}{
   \begin{list}{$\bullet$}
    { \setlength{\itemsep}{0pt}      \setlength{\parsep}{3pt}
      \setlength{\topsep}{3pt}       \setlength{\partopsep}{0pt}
      \setlength{\leftmargin}{3.5mm} \setlength{\labelwidth}{1em}
      \setlength{\labelsep}{0.5em} }
}

\newcommand{\squishend}{
    \end{list}  }

\def\BibTeX{{\rm B\kern-.05em{\sc i\kern-.025em b}\kern-.08em
    T\kern-.1667em\lower.7ex\hbox{E}\kern-.125emX}}

\usepackage{enumitem}
\usepackage{tcolorbox}

\usepackage{xcolor}
\usepackage{listings}

\definecolor{codegray}{rgb}{0.5,0.5,0.5}
\definecolor{codeblue}{rgb}{0.1,0.1,0.8}
\definecolor{backcolour}{rgb}{0.98,0.98,0.98}

\lstdefinestyle{donglinisstaStyle}{
    backgroundcolor=\color{backcolour},   
    commentstyle=\color{codegray}\itshape,
    keywordstyle=\color{codeblue}\bfseries,
    numberstyle=\tiny\color{codegray},
    stringstyle=\color{orange},
    basicstyle=\ttfamily\footnotesize, 
    breakatwhitespace=false,         
    breaklines=true,                 
    captionpos=b,                    
    keepspaces=true,                 
    numbers=left,                    
    numbersep=5pt,                  
    showspaces=false,                
    showstringspaces=false,
    showtabs=false,                  
    tabsize=2,
    frame=single,                    
    rulecolor=\color{gray!30},
    xleftmargin=1.5em,               
}

\lstdefinelanguage{JavaScript}{
  keywords={break, case, catch, continue, debugger, default, delete, do, else, false, finally, for, function, if, in, instanceof, new, null, return, switch, this, throw, true, try, typeof, var, void, while, with, const, let, yield, async, await},
  morekeywords={test, setVariable, setPosition, broadcastMessage, wait, assert, getVariable, isHidden}, 
  sensitive=true,
  morecomment=[l]{//},
  morecomment=[s]{/*}{*/},
  morestring=[b]",
  morestring=[b]',
  morestring=[b]`
}

\usepackage{simplekv}
\usepackage{scratch3}
\usepackage{algorithm}
\usepackage{algpseudocode}
\usepackage{amsmath}

\tcbuselibrary{skins}

\author{Donglin Li}
\affiliation{%
  \institution{Anhui University of Science and Technology}
  \city{Huainan}
  \country{China}
  }
\email{2023301071@aust.edu.cn}

\author{Daming Li}
\affiliation{%
  \institution{Independent Researcher}
  \country{USA}
  }
\email{damingliyale22@gmail.com}

\author{Hanyuan Shi}
\affiliation{%
  \institution{Independent Researcher}
  \country{China}
  }
\email{shihanyuan1995@gmail.com}

\author{Jialu Zhang}
\authornote{Corresponding Author}
\affiliation{%
  \institution{University of Waterloo}
  \city{Waterloo}
  \country{Canada}
  }
\email{jialu.zhang@uwaterloo.ca}

\begin{document}

\title{Raven: Rethinking Automated Assessment for Scratch Programs via Video-Grounded Evaluation}

\begin{abstract}
	Block-based programming environments such as Scratch are widely used in introductory computing education, yet scalable and reliable automated assessment remains elusive. Scratch programs are highly heterogeneous, event-driven, and visually grounded, which makes traditional assertion-based or test-based grading brittle and difficult to scale. As a result, assessment in real Scratch classrooms still relies heavily on manual inspection and delayed feedback, introducing inconsistency across instructors and limiting scalability.

We present \app, an automated assessment framework for Scratch that replaces program-specific state assertions with instructor-specified, task-level video generation rules shared across all student submissions. \app integrates large language models with video analysis to evaluate whether a program’s observed visual and interactive behaviors satisfy grading criteria, without requiring explicit test cases or predefined outputs. This design enables consistent evaluation despite substantial diversity in implementation strategies and interaction sequences.

We evaluate \app on 13 real Scratch assignments comprising over 140 student submissions with ground-truth labels from human graders. The results show that \app significantly outperforms prior automated assessment tools in both grading accuracy and robustness across diverse programming styles. A classroom study with 30 students and 10 instructors further demonstrates strong user acceptance and practical applicability. Together, these findings highlight the effectiveness of task-level behavioral abstractions for scalable assessment of open-ended, event-driven programs.
\end{abstract}

\maketitle

\section{Introduction}
\label{sec:intro}

Scratch~\cite{maloney2010scratch} is a block-based programming language that has become a foundational platform for introductory programming education, with over 100 million learners worldwide~\cite{scratchStats2026}. Its event-driven, visual programming model enables novices to create interactive animations and games, but also poses unique challenges for automated assessment. Manual grading provides rich feedback but is labor-intensive and, as Moreno-Le{\'o}n et al.\ observe, becomes ``\emph{almost impossible in large-scale settings}''~\cite{moreno2015dr}. Educators must either set fewer assessment tasks or resign themselves to a greatly increased marking load~\cite{howshallweassess03}, making human-centric evaluation increasingly unsustainable~\cite{Ala-Mutka01062005}.

Beyond scalability, a more fundamental challenge lies in assessment validity. As the Scratch creator Mitchel Resnick argues, current automated methods often fall into the trap of measuring ``\emph{what is easy to measure, rather than what is important}''~\cite{resnick2017lifelong}. Static analysis fails to capture the dynamic, interactive nature of creative coding, obscuring the actual learning process. While the program code is visible, the visual and interactive execution behavior—the primary learning artifact in Scratch—often remains unexamined. Thus, there remains a critical dearth of valid tools~\cite{Autotoolsysreview24} capable of bridging the gap between static code structure and dynamic visual execution.

In recent years, several automated assessment approaches have been proposed for Scratch programs, among which the \textsc{Whisker} framework~\cite{stahlbauer2019whisker} represents one of the most advanced efforts to date. \textsc{Whisker} validates program behavior by simulating user interactions and checking state assertions. However, its reliance on precise, manually specified state assertions makes it difficult to accommodate the diverse implementation strategies and inherently visual evaluation requirements commonly observed in Scratch projects. As a result, existing tools have not fundamentally resolved the assessment bottleneck faced by large-scale, real-world programming education.

We argue that the core limitation of prior approaches is the lack of a \emph{task-level abstraction} for specifying and evaluating interactive behavior. In practice, students implement the same task using widely varying control flows, event structures, and interaction patterns, yet assessment criteria are defined at the level of the task, not individual programs. Program-specific assertions and hard-coded execution assumptions therefore fail to generalize across submissions.

To address this gap, we propose \app, a video-grounded framework that \textbf{r}ethinks automated \textbf{a}ssessment for Scratch programs via \textbf{v}ideo-based \textbf{e}valuation of program executio\textbf{n}. \app evaluates submissions at the task level using shared video-generation rules that specify how programs are exercised. Instead of relying on program-specific state assertions, it grades by analyzing execution videos. This design decouples grading criteria from individual implementations, enabling consistent assessment across diverse student solutions.

The key insight of \app is to integrate LLMs with video analysis to assess whether a program’s observed visual and interactive behavior satisfies grading criteria. LLMs enable semantic interpretation of high-level visual outcomes, such as geometric correctness or interaction responses, that are difficult to express as symbolic assertions or predefined outputs.

For example, in an assignment requiring students to draw a five-pointed star, assertion-based tools struggle to verify geometric correctness or detect runtime errors such as partial drawings outside the stage. In contrast, these properties are readily observable in execution videos. By analyzing such videos, \app can correctly assess visually valid stars regardless of drawing strategy or code path. Moreover, \app incorporates a lightweight video understanding module that enables automated responses to interactive \texttt{ask} blocks, allowing it to evaluate programs with complex human--computer interaction behaviors that defeat prior tools.

The main contributions of this paper are as follows:

\begin{enumerate}
    \item We present \app, the first automated assessment framework for Scratch that integrates LLMs with video-based execution analysis, enabling robust grading of visually grounded, interactive programs beyond traditional state-assertion-based approaches.

    \item We introduce an assignment-level video generation abstraction that allows instructors to specify assessment criteria via shared interaction sequences, ensuring consistent and scalable evaluation across diverse student implementations.

    \item We design a lightweight video understanding module that supports human--computer interaction patterns in Scratch, including programs that use \texttt{ask} blocks.

    \item We evaluate \app on 13 real-world assignments with over 140 student submissions, and a live classroom study with 30 students and 10 instructors, demonstrating substantial improvements in grading accuracy and instructor acceptance over prior tools.
\end{enumerate}

\section{Background}
\label{sec:background}

\begin{figure}[t!]
    \centering
    \includegraphics[width=1\textwidth]{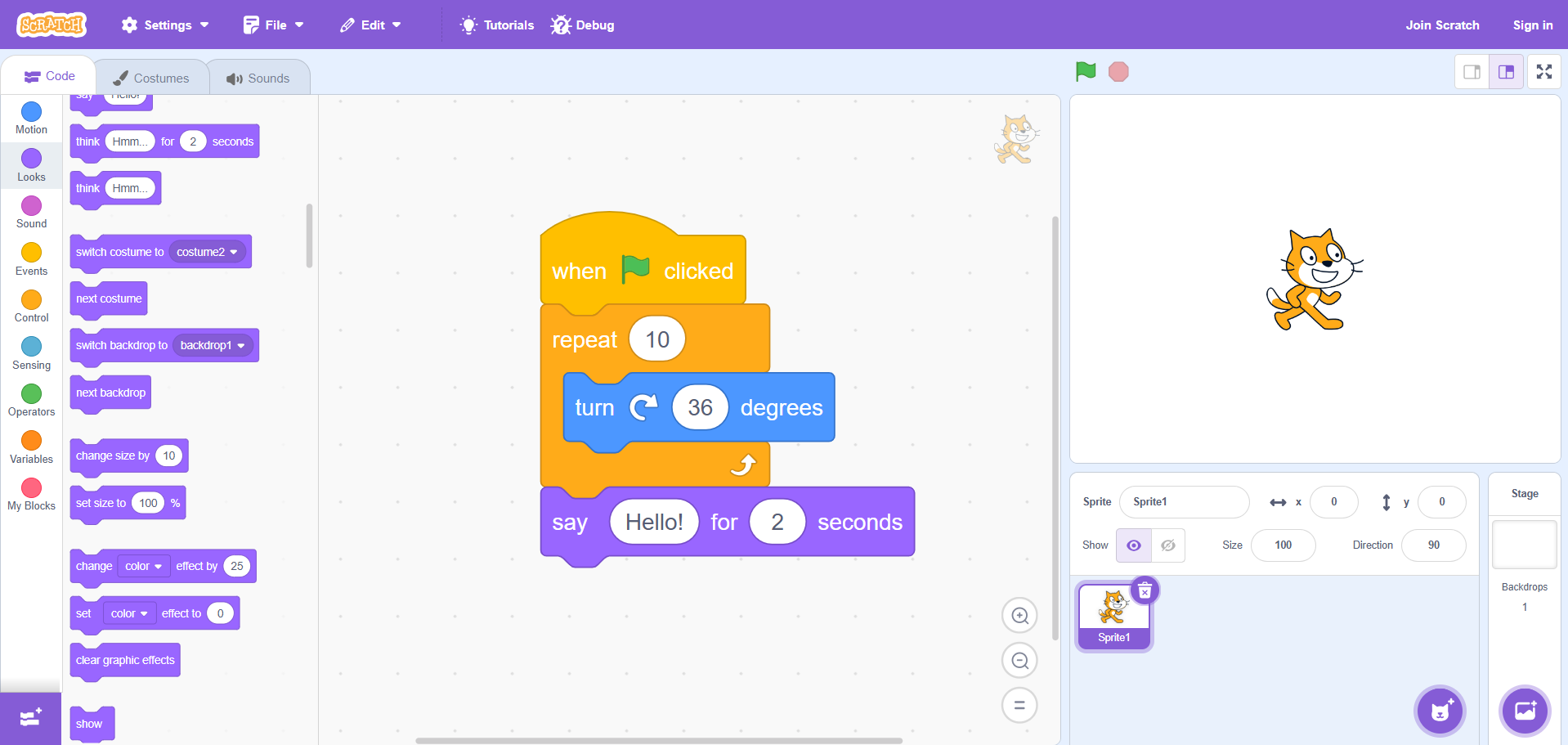}
    \caption{An example Scratch project in which a cat sprite turns around to say hello.}
    \label{fig:scratch_intro}
\end{figure}

Scratch is a visual programming language designed for young and novice programmers~\cite{resnick2009scratch}. It enables users to create interactive stories, games, and animations by dragging and snapping together colored programming blocks.

As illustrated in Fig.~\ref{fig:scratch_intro}, the Scratch programming environment consists of four main regions: (1) the \emph{block palette}, located on the left, which provides available programming blocks; (2) the \emph{code area}, located in the center, where blocks are composed into scripts; (3) the \emph{stage}, located in the upper-right corner, which displays the program’s runtime behavior; and (4) the \emph{sprite list}, located in the lower-right corner, which manages all sprites and backgrounds in a project.

Unlike traditional textual programs, a Scratch project is composed of multiple scripts that execute concurrently. Each script is triggered by an event block (e.g., \emph{when green flag clicked}) and controls sprite behavior through motion, control, and appearance blocks. As a result, Scratch programs naturally exhibit event-driven and concurrent execution behavior.

As computational thinking education has expanded globally, Scratch has become the most widely adopted block-based programming platform, reaching over 100 million learners worldwide~\cite{scratchStats2026}. This widespread adoption has motivated research on automated techniques for analyzing and assessing Scratch programs.

\para{Automated Testing: State-of-the-art Automatic Assessment for Scratch Programs.}
Scratch projects differ fundamentally from conventional programs in that they lack a single entry point. Instead, execution is driven by multiple event handlers (e.g., \emph{when green flag clicked}, \emph{when I receive message}), which may be triggered concurrently. Automated assessment of such event-driven programs requires explicitly modeling user interactions and their effects on program execution.

\textsc{Whisker}~\cite{stahlbauer2019whisker} is the state-of-the-art automated testing framework for Scratch. It evaluates program behavior by simulating user interactions---such as key presses, mouse clicks, and message broadcasts---to trigger events, and by checking whether the resulting system state satisfies expected conditions under a specified interaction sequence.

In \textsc{Whisker}, a test case is specified as a JSON-style script consisting of three core components: (1) an \emph{initial state configuration}, which defines background settings, sprite positions, and variable initial values; (2) an \emph{interaction sequence}, which describes how the user interacts with the program (e.g., key presses or mouse clicks); and (3) \emph{assertions}, which verify whether the system state satisfies expected properties (e.g., \texttt{assert(sprite.x == 100)}, \texttt{assert(variable.score == 10)}).

Listing~\ref{lst:whisker-example} shows a simplified \textsc{Whisker} test case for a fruit-catching game.
Such test cases require precise specification of timing, spatial relationships, and program state assertions, imposing a substantial technical burden on instructors. In practice, many Scratch educators struggle to manually design test cases that are both correct and effective. To mitigate this burden, prior work~\cite{DBLP:journals/corr/abs-2009-04115,götz2022modelbasedtestingscratchprograms,deiner2023automated} has explored the automatic generation of \textsc{Whisker} test cases.

\begin{lstlisting}[
    style=donglinisstaStyle, 
    float,
    floatplacement=t,
    language=JavaScript, 
    caption={Example \textsc{Whisker} test case for a fruit-catching game.}, 
    label={lst:whisker-example}
]
test("player catches apple", () => {
  // 1. Initialization
  setVariable("score", 0);
  setPosition("apple", {x: 50, y: 100});
  setPosition("basket", {x: 50, y: -150});

  // 2. Interaction sequence
  broadcastMessage("start_game");
  wait(200);

  // 3. Assertions
  assert(getVariable("score") === 1);
  assert(isHidden("apple") === true);
});
\end{lstlisting}

In this work, we adopt \textsc{Whisker}’s default MIO-based test generation algorithm~\cite{deiner2023automated}, reflecting how automated testing tools are commonly used in practice. Given a generated test suite, \textsc{Whisker} evaluates a student program by initializing the execution state, replaying the interaction sequences, and checking assertions against observed runtime states, ultimately producing a pass/fail verdict.

\section{Motivation and Technical Challenges}
\label{sec:motivation}

Despite substantial progress in automated assessment for Scratch, existing tools remain fundamentally limited in real-world educational settings. Scratch programs are open-ended, visually oriented, and highly heterogeneous, making assessment approaches based on rigid execution assumptions difficult to generalize across authentic student submissions. As the Scratch creator Mitchel Resnick emphasizes, ``\emph{languages need `wide walls' (supporting many different types of projects so people with many different interests and learning styles can all become engaged)}''~\cite{resnick2009scratch}. However, when deployed in classrooms, current automated assessment systems often impose restrictive assumptions that conflict with instructional practice, leading to brittle evaluations and unreliable grading outcomes. We identify four recurring technical challenges that expose this mismatch.

\begin{figure}
    \centering
    \includegraphics[width=1\linewidth]{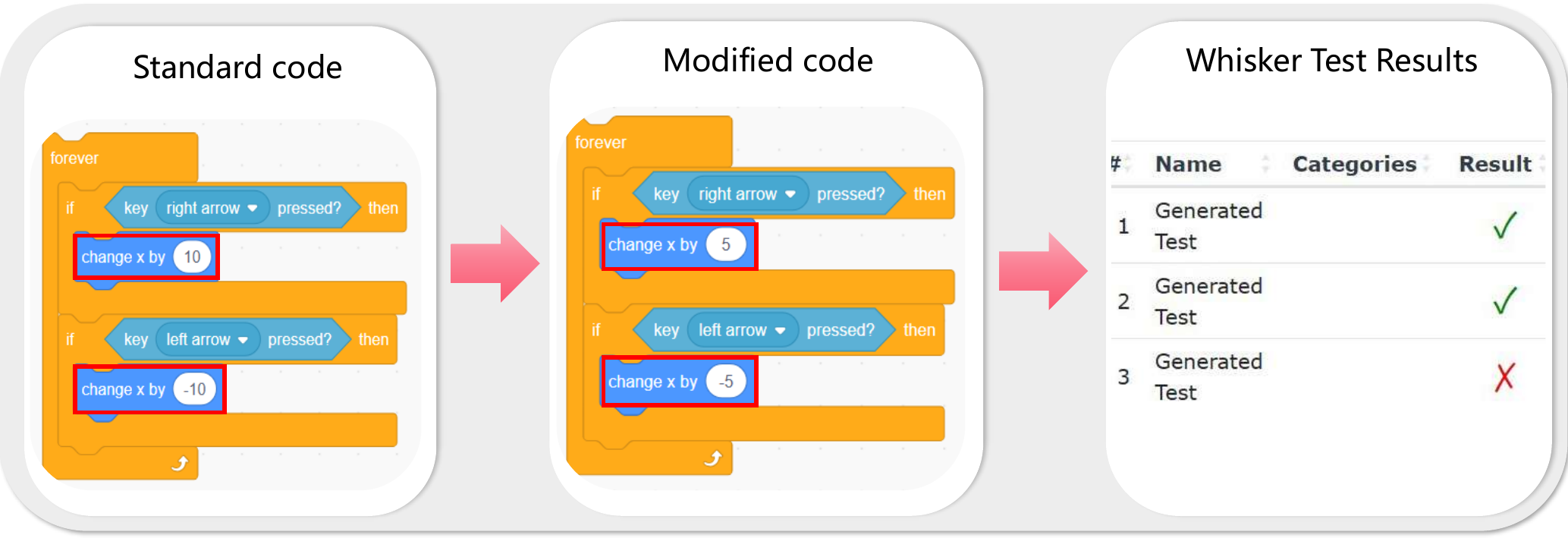}
    \caption{Reducing the number of steps taken from 10 to 5 is a reasonable change in real-world evaluations. However, \textsc{Whisker} yields incorrect evaluation results.}
    \label{fig:Challenge1}
\end{figure}

\para{Challenge 1: Inadequate Coverage of Open-ended Task Requirements.}
In real classrooms, a single Scratch assignment often admits multiple correct implementation strategies. Even within the same strategy, students may legitimately vary parameters such as movement speed, timing, or intermediate states while still satisfying instructional goals.

For example, consider an assignment requiring a sprite to move left and right in response to arrow key presses (Fig.~\ref{fig:Challenge1}). A reference solution may update the sprite’s $x$-coordinate by 10 units per key press, whereas a student implementation may move the sprite by 5 units per press. Although the latter produces slower motion, it exhibits the same intended visual behavior and meets the task requirements from an instructional perspective. Nevertheless, assertion-based testing in \textsc{Whisker} flags such submissions as incorrect.

This failure arises because \textsc{Whisker} relies on precise state assertions derived from a reference implementation. Small deviations in parameter choices alter subsequent execution states, leading to cascading assertion violations and incorrect grading outcomes, despite functionally acceptable observable behavior.

\begin{figure}
    \centering
    \includegraphics[width=0.9\linewidth]{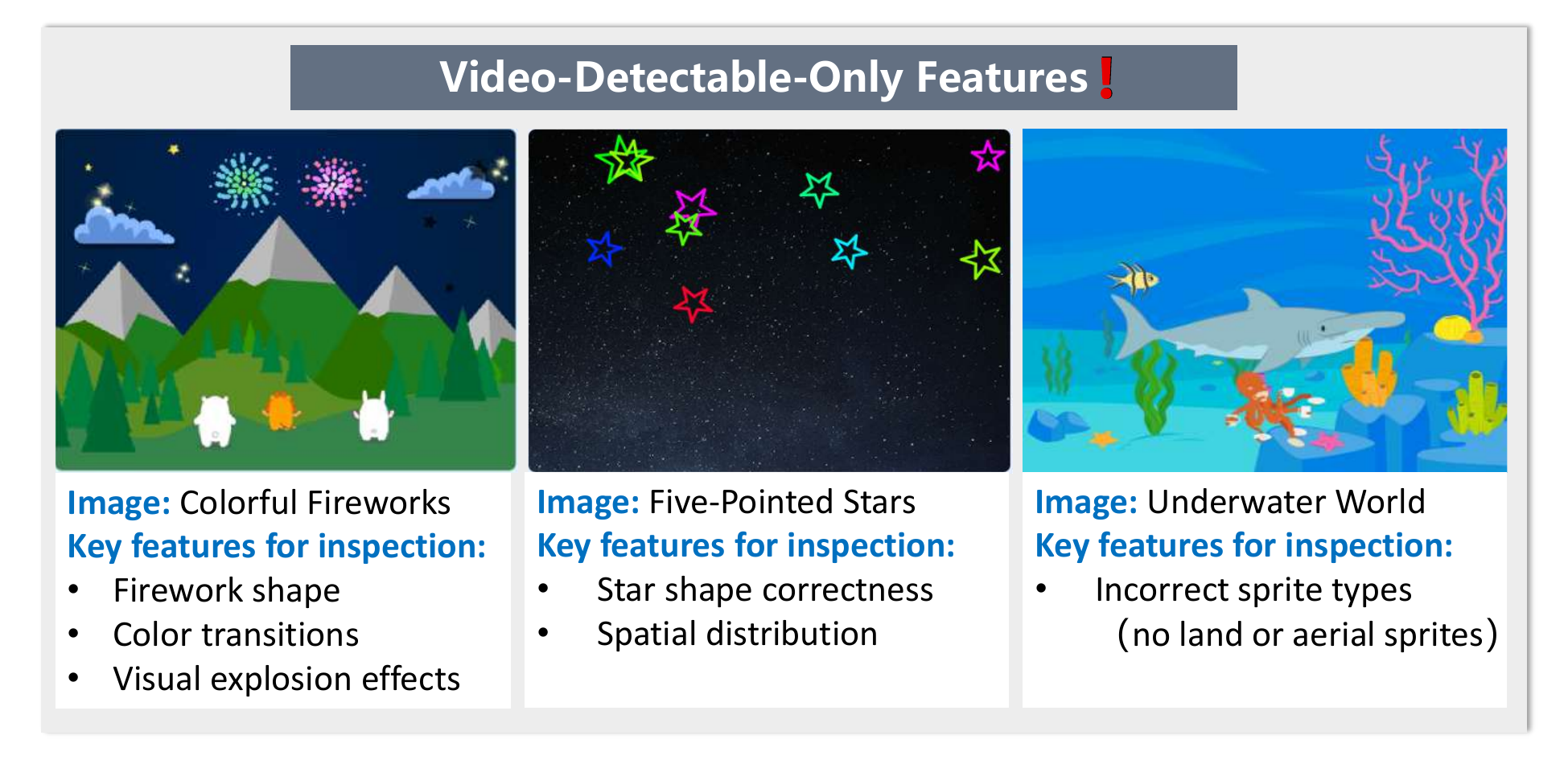}
    \caption{Visual features -- such as firework shapes, color transitions, and spatial distribution -- can only be verified via video, whereas assertion-based tools like \textsc{Whisker} struggle to capture or evaluate them.}
    \label{fig:Challenge2}
\end{figure}

\para{Challenge 2: Inability to Assess Visually Defined Correctness.}
Many Scratch assignments define correctness primarily in terms of visual outcomes rather than program state. As illustrated in Fig.~\ref{fig:Challenge2}, examples include requiring that an underwater scene contains no land or aerial sprites, that fireworks visually explode in the sky, or that a program draws multiple five-pointed stars with specific visual properties.
Such requirements are difficult to formalize using state assertions alone. In practice, evaluating these assignments requires executing the Scratch program and judging its rendered animation. Assertion-based tools such as \textsc{Whisker}, which reason over predefined variables and sprite states, are therefore ill-suited to assess visually grounded correctness criteria.

\begin{figure}
    \centering
    \includegraphics[width=0.6\linewidth]{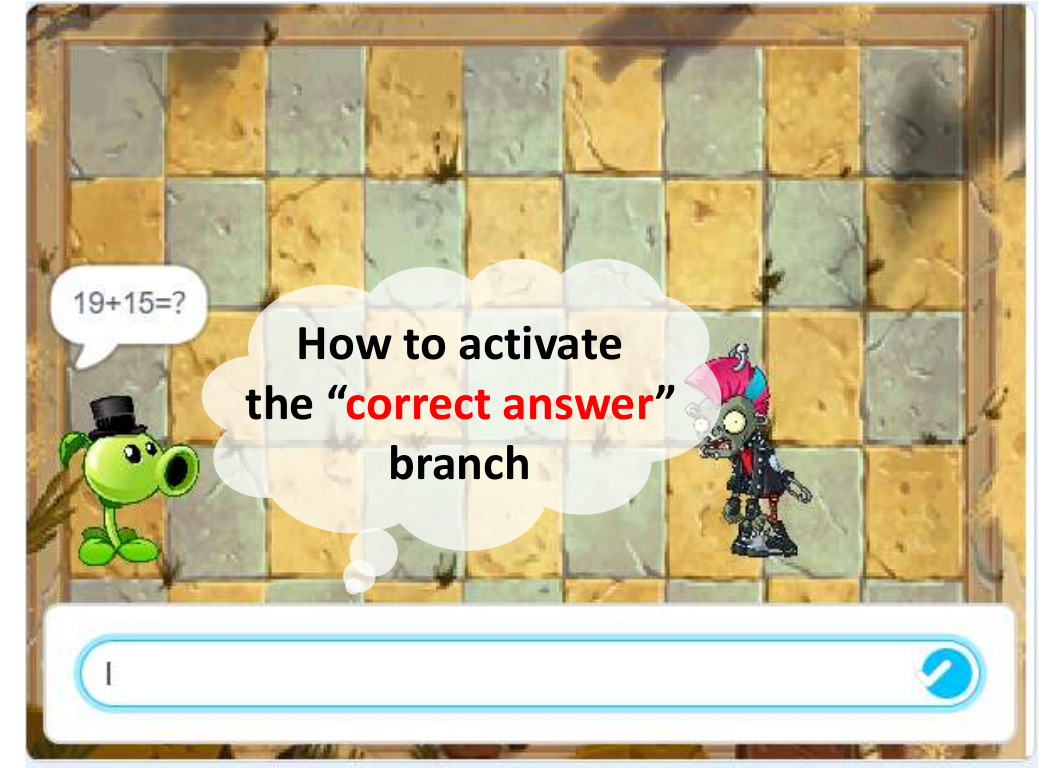}
    \caption{In a ``Math Pea Shooter'' game, \textsc{Whisker} fails to provide the correct inputs required to trigger the specific logic for shooting zombies with shells.}
    \label{fig:Challenge3}
\end{figure}

\para{Challenge 3: Limited Support for Human--computer Interaction Via \texttt{ask} Blocks.}
Human--computer interaction is a core component of Scratch programming. The \texttt{ask} block enables sprites or backgrounds to prompt users for input and conditionally alter behavior based on the response, and is often the only mechanism for implementing interactive logic in Scratch projects.
However, \textsc{Whisker} provides limited strategies for handling \texttt{ask} blocks, typically relying on random inputs or simple heuristics when the input is guarded by specific conditional statements. 
For example, as illustrated in Fig.~\ref{fig:Challenge3}, in a ``Math Pea Shooter'' game, the player must correctly answer an arithmetic question to trigger a projectile attack. Despite repeated attempts using automatically generated \textsc{Whisker} test cases, the tool consistently fails to activate the ``correct answer'' branch, rendering automated assessment ineffective.

\begin{figure}
    \centering
    \includegraphics[width=0.8\linewidth]{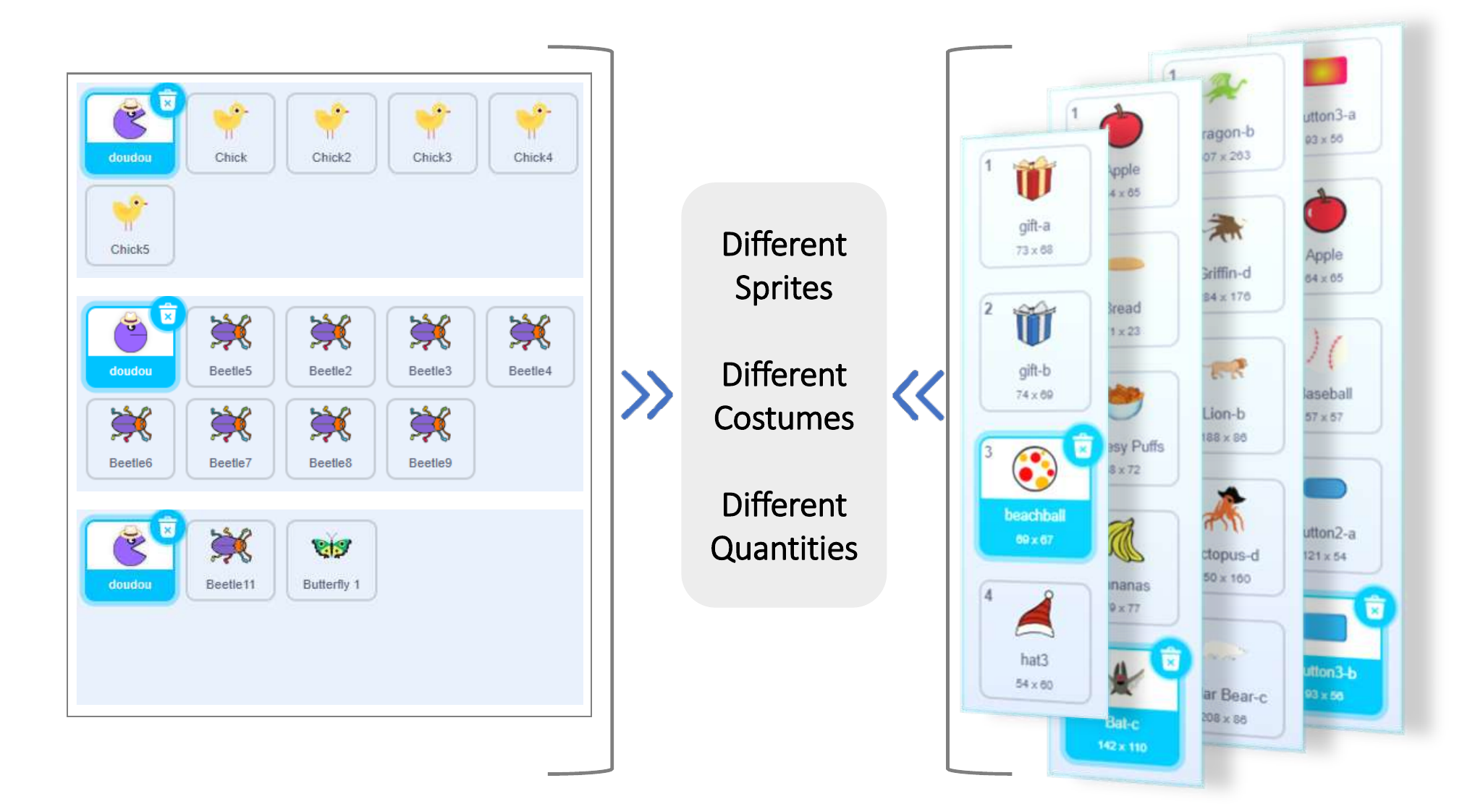}
    \caption{In real classroom settings, many Scratch assignments allow flexibility in the choices, costumes, and quantities of sprites.}
    \label{fig:Challenge4}
\end{figure}

\para{Challenge 4: Fragility Under Flexible Sprite and Background Configurations.}
In real classroom settings, many Scratch assignments allow flexibility in the number, names, and visual appearances of sprites and backgrounds. In our dataset, a substantial fraction of assignments permit such variations. However, \textsc{Whisker} assumes fixed sprite identities and configurations derived from a reference program.
As a result, when students choose different sprite names, costumes, or quantities, assertion-based automated tests fail to locate the expected entities. As illustrated in Fig.~\ref{fig:Challenge4}, in a ``Magic Show'' game (right), students use various costumes to represent magic props, leading to mismatches in sprite names and appearances. Similarly, in a ``Pea Shooter (Pests)'' game (left), students use varying numbers and types of pest sprites, causing test failures due to unrecognized sprite identities, even when the programs meet the task requirements.

Together, these challenges highlight a fundamental mismatch between assertion-based automated testing and the realities of open-ended, visually driven, and interaction-rich Scratch programming. Addressing these limitations requires rethinking the abstraction underlying automated assessment, rather than merely improving test generation techniques.

\begin{table}[t!]
\centering
\small
\begin{tabular}{p{3.0cm} p{1.5cm} p{1.6cm} l}
\toprule
\textbf{Assignment} &
\textbf{{LDE}} &
\textbf{{VDOE}} &
\textbf{Examples of Video-Detectable Errors} \\
\midrule
Underwater World & 50\% & 50\% & Incorrect sprite types (land/air animals) \\
Colorful Fireworks & 80\% & 20\% & Firework shape, color transitions, explosion effect \\
Magic Show & 0\% & 100\% & Object switching timing, hat open/close state \\
Draw Five-Pointed Star & 71\% & 29\% & Star shape correctness, spatial distribution \\
Ball Goes Up & 50\% & 50\% & Premature motion, incorrect movement direction \\
Big Fish Eats Small Fish & 50\% & 50\% & Facing direction, smoothness, respawn animation \\
Pea Shooter & 0\% & 100\% & Enemy animation, hit feedback, score visualization \\
Thinking Graphics & 100\% & 0\% & Dice face shape and number consistency \\
Catch the Eggs & 78\% & 22\% & Falling animation, catch feedback, scoring effect \\
Hit the Bats & 100\% & 0\% & Random bat movement, hit animation feedback \\
Mental Math Race & 44\% & 56\% & Answer feedback, finish-time visualization \\
Math Pea Shooter & 0\% & 100\% & Projectile direction, hit animation, defeat effects \\
Rock--Paper--Scissors & 20\% & 80\% & Hover/click animations, sprite-choice alignment \\
\bottomrule
\end{tabular}
\caption{Comparison of Error Detectability: Logic-Detectable vs. Video-Detectable-Only. We categorize errors into two classes.
(1) Logic-detectable errors (\textbf{LDE}), which can be identified through static or logic-based analysis; and
(2) Video-detectable-only errors (\textbf{VDOE}), which cannot be detected by logic analysis alone and require execution videos for reliable identification.
If an error is detectable by both logic-based and video-based approaches, we classify it as logic-detectable, reserving the ``video-only'' category strictly for cases where video evidence is essential.
We intentionally reserve the term video-detectable-only for errors whose detection fundamentally depends on visual execution evidence.}
\label{tab:logic-vs-video}
\end{table}

\para{Why Video Input Is Crucial for Scratch Program Assessment?}
Traditional Scratch assessment tools determine correctness by reasoning over program logic and state assertions. This paradigm assumes that correct behaviors can be specified by a fixed interaction sequence and precise system states—an assumption that rarely holds in real Scratch classrooms. In assertion-based frameworks such as \textsc{Whisker}, test cases must strictly follow predefined interaction sequences, and all asserted states must exactly match reference values. Even minor deviations in interaction order, timing, movement granularity, or sprite configuration can alter execution paths and trigger cascading assertion failures, causing functionally correct programs to be incorrectly labeled as wrong.

\app adopts a fundamentally different approach by evaluating correctness from execution videos. Given instructor-defined video generation rules, the system infers whether the observed visual behavior satisfies grading criteria. As a result, variation in interaction sequences, implementation strategies, sprite choices, and intermediate states does not affect grading outcomes, as long as the rendered animation meets task requirements. This shift is crucial because many Scratch assignments define correctness in terms of visual outcomes rather than internal logical states (e.g., whether fireworks visually explode or a star is drawn with the correct shape)—properties that are difficult or impossible to capture using static analysis or state assertions alone. An analysis of 13 representative Scratch assignments confirms that a substantial fraction of incorrect submissions exhibit errors detectable only through visual inspection of execution videos (Table~\ref{tab:logic-vs-video}). These findings show that effective automated assessment for Scratch must treat runtime visual behavior as a first-class artifact, as embodied by \app.

\begin{figure}
    \centering
    \includegraphics[width=1\linewidth]{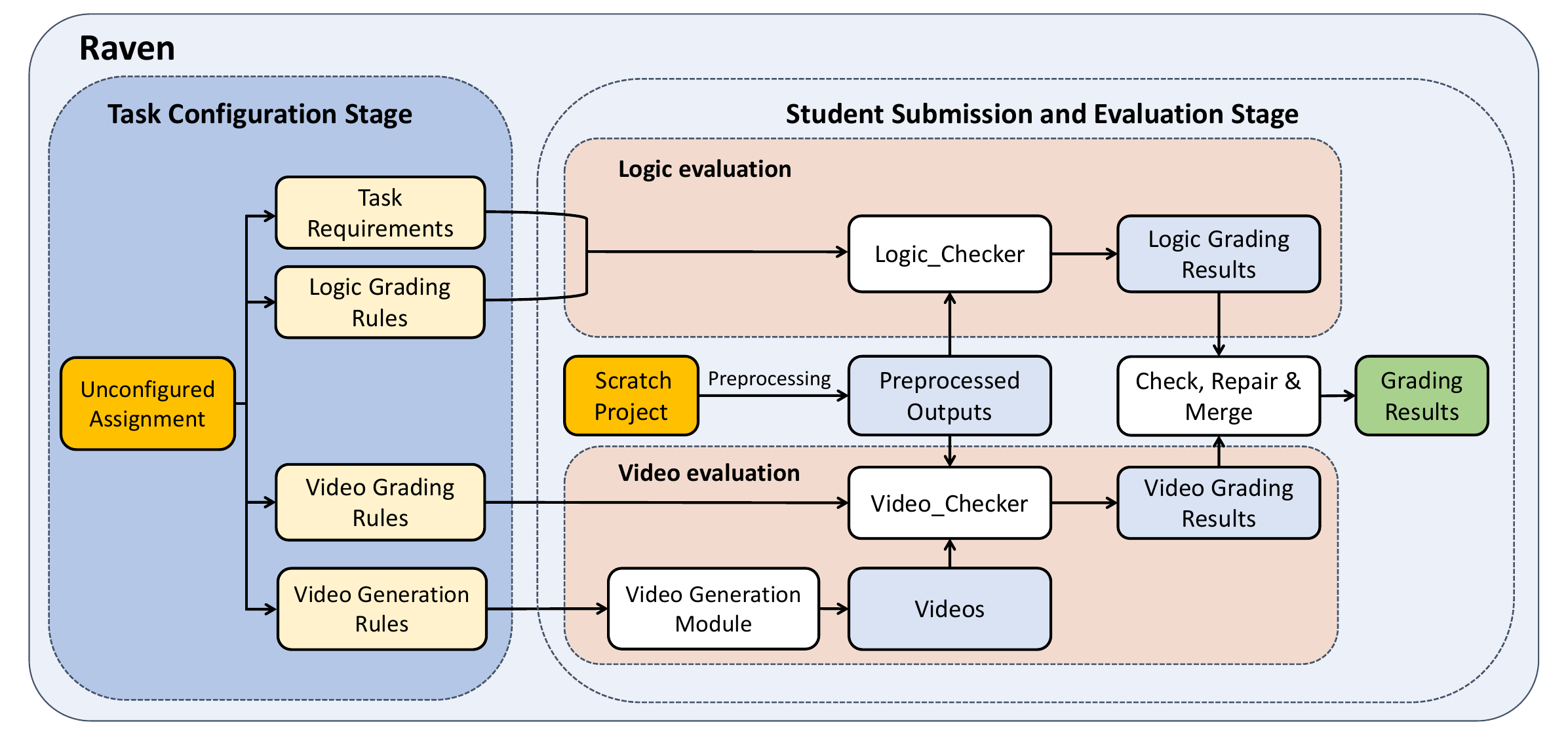}
    \caption{Architecture of the \app Framework. The system evaluates projects via a dual-track pipeline (Logic and Video) organized into two stages: 
(1) \textbf{Task Configuration Stage}: The instructor initializes a \textit{Unconfigured Assignment} (orange-yellow box) to define specific \textit{Task Requirements, Logic Grading Rules and Video Grading Rules} (light yellow boxes). 
(2) \textbf{Student Submission and Evaluation Stage}: A student's \textit{Scratch Project} (orange box) is processed by \textit{Functional Operators} (white boxes, e.g., \texttt{Logic\_Checker}, \texttt{Video Generation Module}, and \texttt{Video\_Checker}). These operators generate \textit{Intermediate Results} (light blue boxes), which are then synthesized into the \textit{Final Grading Results} (green box).}
    \label{fig:system}
\end{figure}

\section{System Overview}
\label{sec:system_design}

\app operates in two stages: a \emph{task configuration stage} and a \emph{student submission and evaluation stage}. This separation enables instructors to flexibly customize assignments and grading criteria, while allowing students' submissions to be evaluated automatically and consistently. Below, we outline the system architecture (Fig.~\ref{fig:system}) and the algorithmic workflow (Algorithm~\ref{alg:Raven}) and define the key terminology.

\begin{algorithm}
\caption{\app: Hybrid Logic-Video Evaluation}
\label{alg:Raven}
\begin{algorithmic}[1]
\Require Student project file $\mathcal{S}$, Task description $\mathcal{Q}$, Grading rules $\mathcal{T} = \{\mathcal{T}_{log}, \mathcal{T}_{vid}\}$, Video generation rules $\mathcal{G}$
\Ensure Final comprehensive grading report $\mathcal{R}_{final}$

\Statex \textbf{// Phase 1: Preprocessing}
\State $\mathcal{D} \leftarrow \textsc{Preprocess}(\mathcal{S})$ \Comment{Extract project JSON and initial states}
\Statex \textbf{// Phase 2: Logic Evaluation}
\State $\mathcal{R}_{log} \leftarrow \textsc{Logic\_Checker}(\mathcal{Q}, \mathcal{T}_{log}, \mathcal{D})$ \Comment{Static analysis based on task description}

\Statex \textbf{// Phase 3: Video Evaluation}
\State $\mathcal{V} \leftarrow \textsc{GenerateVideos}(\mathcal{S}, \mathcal{G})$ \Comment{Record videos based on distinct video generation rules}

\For{\textbf{each} $v \in \mathcal{V}$} \Comment{Iterate through each generated test-case video}
    \For{$run \gets 1$ \textbf{to} $3$} \Comment{Mitigating VLM stochasticity for the current video}
        \State $\mathcal{F} \leftarrow \textsc{ExtractFrames}(v)$ \Comment{Sampling and encoding key frames from $v$}
        \State $\mathcal{R}_{tmp}[run] \leftarrow \textsc{Video\_Checker}(\mathcal{F}, \mathcal{T}_{vid}, \mathcal{D})$ \Comment{Visual inspection via VLM}
    \EndFor
    \State $\mathcal{R}_{vid} \leftarrow \textsc{MergeRuns}(\mathcal{R}_{vid}, \{\mathcal{R}_{tmp}\})$ \Comment{Update $\mathcal{R}_{vid}$ using a lower-bound aggregation}
\EndFor

\Statex \textbf{// Phase 4: Result Checking and Merging}
\State $\mathcal{R}_{final} \leftarrow \textsc{Check\_Repair\_Merge}(\mathcal{R}_{log}, \mathcal{R}_{vid})$ \Comment{Reconcile logic and video results}
\State \Return $\mathcal{R}_{final}$
\end{algorithmic}
\end{algorithm}

\subsection{Task Configuration Stage}
\label{task_config}
In the task configuration stage, instructors define the assignment by providing four components: (1) a task description, (2) logic grading rules (3) video grading rules, and (4) video generation rules. Hereafter, (2) and (3) are collectively referred to as \textit{grading rules}. This stage offers substantial flexibility, allowing instructors to tailor assignments to different instructional goals, student skill levels, and assessment scenarios.

\para{Task Description.} The \emph{task description} specifies the creative requirements presented to students and describes the intended behavior of the Scratch project.  

\begin{figure}
\centering
\begin{tcolorbox}[
    colback=gray!5,
    colframe=gray!75,
    left=5pt, right=5pt, top=5pt, bottom=5pt,
    arc=0mm,
    boxrule=0.5pt,
    enhanced jigsaw
]
\ttfamily\footnotesize\raggedright
\setlist[enumerate]{leftmargin=1.5em, nosep, labelsep=0.5em}
\setlist[itemize]{leftmargin=1.5em, nosep, labelsep=0.5em}

You are a professional programming education assessment expert, specializing in logic-level evaluation of Scratch projects.

You are able to precisely compare differences between a student's implementation and the expected objectives, and provide fair and impartial scores and feedback.

{\color[RGB]{204, 0, 0}
You must first obtain the necessary information through tools before answering. Do not guess or make assumptions.

You are required to strictly follow the workflow below} to generate a grading report:

\begin{enumerate}[label=\arabic*.]
    \item 
    Read the task description and logic grading rules below, and learn the task requirements and the logic grading criteria.
    
    \colorbox[RGB]{240, 240, 240}{\mbox{\hspace{2pt}\textit{<Task Description> + <Logic Grading Rules>}\hspace{2pt}}}

    \item 
    Logic evaluation: {\color[RGB]{204, 0, 0}
    Strictly follow the grading criteria and perform the evaluation as follows:}
    \begin{itemize}[label={-}]
        \item 
        Read and understand the Preprocessed Outputs below and the analyzed states of all sprites and backgrounds.
        
        \colorbox[RGB]{240, 240, 240}{\mbox{\hspace{2pt}\textit{<Preprocessed Outputs>}\hspace{2pt}}}
        
        \item 
        For each grading item, perform logic-level verification on the student Scratch project.
        
        \item 
        If the verification meets the expected requirement, the grading item receives full marks; otherwise, it receives 0 points.
        
        \item 
        List the score for each logic grading item in the {\color[RGB]{49, 130, 189}following format:}
        \begin{itemize}[label={-}]
            \item {\color[RGB]{49, 130, 189}Logic Grading Rule 1: [Grading Rule Description] + [Scoring Evidence (must describe in detail: which code blocks or states were verified and whether all conditions were met)] + Scoring Result (xx points)}
            \item {\color[RGB]{49, 130, 189}Logic Grading Rule 2: [Grading Rule Description] + [Scoring Evidence] + Scoring Result (xx points)}
            \item {\color[RGB]{49, 130, 189}Logic Grading Rule 3: [Grading Rule Description] + [Scoring Evidence] + Scoring Result (xx points)}
            \item {\color[RGB]{49, 130, 189}...}
            \item {\color[RGB]{49, 130, 189}Total Score: xx points}
        \end{itemize}
    \end{itemize}
\end{enumerate}

\vspace{0.5em}

{\color[RGB]{204, 0, 0}
Carefully list the score for each grading item and compute the total score. Please note that your response must not include tool usage details. Only include the final analysis results.

Your response must contain only numbers, punctuation marks and text. Do not include any other content, including emojis, images, tables, code or formatting.}

\end{tcolorbox}
\caption{Prompt for the logic\_checker}
\label{fig:prompt_logic_checker}
\end{figure}

\para{Grading Rules.}
The \emph{grading rules} define how student submissions should be evaluated in a fair and uniform manner. Each rule consists of a natural-language description and an associated score. A submission receives the score only if it fully satisfies the rule; otherwise, it receives zero. Grading rules are divided into two types.  
\emph{Logic grading rules} specify code-level requirements (e.g., ``If the project implements keyboard-controlled movement for the dog sprite, award 10 points''). \app evaluates whether the functionality is implemented and whether the underlying logic is correct.  
\emph{Video grading rules} specify expected runtime behaviors observable in the rendered animation (e.g., ``If the ball disappears and the score increases when it collides with the dog, award 10 points''). These rules focus exclusively on visual outcomes rather than internal program states.

\para{Video Generation Rules.} 
The \emph{video generation rules} describe how \app should interact with the program to produce execution videos for evaluation. Instructors specify one or more event sequences, which the system follows to automatically operate the Scratch program and record videos. Supported events include keyboard and mouse actions, mouse movements, and responses to \texttt{ask} blocks.

\subsection{Student Submission and Evaluation Stage}

After task configuration, students submit their Scratch projects to \app. As detailed in Algorithm~\ref{alg:Raven}, \app executes a Hybrid Logic-Video Evaluation pipeline. This process begins with static logic inspection to verify code structures against task requirements, followed by the video generation based on instructor-specified rules. These videos undergo robust visual analysis using a multi-run strategy to mitigate VLM stochasticity before being synthesized with logic checks to produce the final grading results. A Scratch project is typically provided in \texttt{.sb3} format, which is unpacked to obtain the JSON-based source code and associated media assets.

\para{Preprocessing.}
Since large language models do not automatically extract salient semantic features from Scratch code, \app first performs a preprocessing step. The system parses the project’s JSON source code to extract background and sprite descriptions, initial states, and structural information. These extracted features are packaged together with the original JSON code and serve as shared input to both logic-based and video-based evaluation modules.

\para{Logic Evaluation.}
For logic evaluation, \app feeds the preprocessed outputs, logic grading rules, and task description into a \texttt{logic\_checker}. The logic checker uses the state-of-the-art large language model \texttt{Qwen3-max}~\cite{yang2025qwen3technicalreport} to compare the student implementation against the intended functionality and produces a logic score. Fig.~\ref{fig:prompt_logic_checker} displays the prompt for the \texttt{logic\_checker}.

\para{Video Evaluation.}
Video evaluation addresses the long-standing difficulty of automated testing in Scratch, where high randomness, complex event-triggered behaviors, and diverse implementation strategies make it challenging to define reliable assertions. Instead of reasoning over abstract system states, \app evaluates correctness directly from execution videos.

\begin{figure}[H]
\centering
\begin{tcolorbox}[
    colback=gray!5,
    colframe=gray!75,
    left=5pt, right=5pt, top=5pt, bottom=5pt,
    arc=0mm,
    boxrule=0.5pt,
    enhanced jigsaw
]
\ttfamily\footnotesize\raggedright
\setlist[enumerate]{leftmargin=1.5em, nosep, labelsep=0.5em}
\setlist[itemize]{leftmargin=1.5em, nosep, labelsep=0.5em}

You are a professional programming education assessment expert, specializing in event-based testing and grading of Scratch projects.

You are able to precisely compare differences between a student's implementation and the expected objectives, and provide fair and impartial scores and feedback.

Video Frame List: \colorbox[RGB]{240, 240, 240}{\mbox{\hspace{2pt}\textit{<Video Frame List>}\hspace{2pt}}}

Brief project analysis: \colorbox[RGB]{240, 240, 240}{\mbox{\hspace{2pt}\textit{<Preprocessed Outputs>}\hspace{2pt}}}

You should only focus on the grading rules provided below:

\colorbox[RGB]{240, 240, 240}{\mbox{\hspace{2pt}\textit{<Video Grading Rules>}\hspace{2pt}}}

{\color[RGB]{204, 0, 0}
Do not guess or make assumptions. You are required to strictly follow the event testing criteria} to generate the evaluation results:

\begin{itemize}[label={-}]
    \item 
    For each grading rule, if the behaviors and events in the video fully comply with all conditions of the criteria, the item receives full marks; otherwise, it receives 0 points.
    
    \item 
    Summarize the evaluation results in the {\color[RGB]{49, 130, 189}following format:}
    \begin{itemize}[label={-}]
        \item {\color[RGB]{49, 130, 189}Grading Rule 1: [Grading Rule Description] + [Scoring Evidence (must describe in detail: which frames were checked, what specific phenomena were observed, and whether all conditions were met)] + Scoring Result (xx points / Full points)}
        \item {\color[RGB]{49, 130, 189}Grading Rule 2: [Grading Rule Description] + [Scoring Evidence] + Scoring Result (xx points / Full points)}
        \item {\color[RGB]{49, 130, 189}...}
        \item {\color[RGB]{49, 130, 189}Total Score: xx points / Full points}
    \end{itemize}
\end{itemize}

\vspace{0.5em}
{\color[RGB]{204, 0, 0}
Your response must contain only numbers, punctuation marks, and text. Do not include any other content, including emojis, images, tables, code, or special formatting markers.}

\end{tcolorbox}
\caption{Prompt for the video\_checker}
\label{fig:prompt_video_checker}
\end{figure}

The video generation module executes the student project according to the instructor-specified video generation rules and records one or more videos. These videos serve as the sole evidence for video-based grading. To handle interactive \texttt{ask} blocks, where traditional tools rely on random or heuristic responses, \app incorporates a lightweight vision model \texttt{Qwen-vl-plus}~\cite{bai2025qwen25vltechnicalreport} that interprets on-screen content and generates human-like responses, enabling effective human--computer interaction during automated execution.

For each video test case, instructors only need to provide simple arrays specifying event sequences, such as keyboard and mouse actions, mouse movements, and responses to \texttt{ask} blocks. This design allows even instructors without technical testing expertise to configure executable test scenarios.

The \texttt{video\_checker} evaluates whether the recorded videos satisfy the video grading rules using the multimodal video analysis model \texttt{Qwen3-vl-plus}~\cite{bai2025qwen3vltechnicalreport}. For each video, \app performs multiple repeated evaluations to reduce the risk of hallucinations. A grading rule is awarded full credit only if it is satisfied across all repeated evaluations; otherwise, no credit is given.  Fig.~\ref{fig:prompt_video_checker} displays the prompt for the \texttt{video\_checker}.

The video checker is provided with three inputs: (1) preprocessed outputs to support semantic understanding of sprites and interactions, (2) instructor-defined video grading rules to guide attention, and (3) execution videos, which are sampled at 10 FPS to balance recognition accuracy and token efficiency.

\subsection{Result Checking and Merging}
\label{sec:check_repair_merge}

Finally, \app performs a consistency check between logic-based and video-based evaluation results using \texttt{Qwen-plus}~\cite{yang2025qwen3technicalreport}, a large language model balancing capability and cost. If conflicts arise, video-based results take precedence, as they directly reflect observable runtime behavior. The system then merges the two evaluation results and computes the final grade for the student submission.

\section{Evaluation}
\label{sec:eval}

We empirically evaluate \app with respect to the following research questions:

\begin{itemize}
    \item \textbf{RQ1:} Can \app accurately assess the correctness of Scratch programs?
    \item \textbf{RQ2:} Is \app scalable and deployable in real classroom environments?
\end{itemize}

\subsection{RQ1: Correctness of Automated Assessment}

\para{Dataset and Experimental Setup.}
We collect data from 13 Scratch assignments used in real-world classroom instruction at an after-school education center, covering multiple difficulty levels, grade ranges, and instructional topics. In total, the dataset contains 146 student submissions. Table~\ref{tab:dataset-complexity} summarizes the characteristics of each assignment, including difficulty level, number of blocks (B), number of scripts (S), inter-procedural cyclomatic complexity (ICC), number of student submissions, and the core Scratch blocks involved.

\begin{table}
\centering
\small
\begin{tabular}{lcccccc}
\toprule
\textbf{Assignment} &
\textbf{Difficulty} &
\textbf{B} &
\textbf{S} &
\textbf{ICC} &
\textbf{\#Submissions} &
\textbf{Core Blocks} \\
\midrule
Underwater World & Easy & 21 & 3 & 7 & 10 & looks, loops, motion \\
Colorful Fireworks & Easy & 40 & 1 & 6 & 10 & special effects, sound, loops \\
Magic Show & Easy & 24 & 3 & 11 & 11 & next costume, events, sound \\
Draw Five-Pointed Star & Easy & 22 & 1 & 4 & 10 & pen, loops, randomness \\
Ball Goes Up & Easy & 41 & 2 & 3 & 12 & looks, motion, events \\
Big Fish Eats Small Fish & Medium & 34 & 2 & 12 & 18 & looks, events, motion, collisions \\
Pea Shooter (Pests) & Medium & 197 & 11 & 44 & 12 & looks, wait, randomness, collisions \\
Thinking Graphics & Medium & 43 & 4 & 10 & 6 & interaction, control, operators \\
Catch the Eggs & Medium & 65 & 2 & 18 & 11 & events, clone, collisions, timing \\
Hit the Bats & Hard & 139 & 5 & 36 & 9 & events, randomness, timing \\
Mental Math Race & Hard & 85 & 5 & 30 & 13 & interaction, timing, variables \\
Math Pea Shooter & Hard & 50 & 3 & 13 & 11 & interaction, collisions, broadcast \\
Rock--Paper--Scissors & Hard & 82 & 7 & 21 & 13 & variables, randomness, broadcast \\
\bottomrule
\end{tabular}
\caption{Characteristics of the Scratch assignments used in evaluation.
B = number of blocks; S = number of scripts; ICC = inter-procedural cyclomatic complexity. Core Blocks showcase the core components in each Scratch assignment.}
\label{tab:dataset-complexity}
\end{table}

To evaluate the grading quality of \app, we recruited five Scratch instructors, each with over two years of Scratch teaching experience, to independently grade the same set of student submissions using identical task descriptions and grading rules. The full score for each assignment is 100, with raw scores given based on grading rules described in section \ref{task_config}.

We define \emph{binary correctness} in such a way that a submission is labeled as \emph{correct} only if the majority of the five instructors assigns a full score of 100, and \emph{incorrect} otherwise. This serves as the ground truth for evaluating the grading quality of \textsc{Whisker} and \app.

Next, we use the reference solution of each assignment as the seed program for \textsc{Whisker} and apply its default test generation mechanism to produce test cases. These test cases are then executed on the remaining student submissions. In parallel, \app evaluates the same submissions using instructor-provided assignment descriptions, grading rules, and video generation rules. 
We report accuracy, precision, recall, and F1 score for both tools.
Accuracy ($Acc$) is defined as the fraction of submissions whose grades
match the ground truth labels.
Precision ($Prec$) is defined as the fraction of submissions graded as
incorrect by the tool that are truly incorrect.
Recall ($Rec$) is defined as the fraction of truly incorrect submissions
that are correctly identified by the tool.
The F1 score is the harmonic mean of precision and recall.


\para{Binary Correctness Grading Results: \app vs. \textsc{Whisker}.}
Table~\ref{tab:binary-results} summarizes the results across all assignments from the binary correctness perspective. Across all the 13 assignments, \app achieves near-perfect performance, significantly outperforming \textsc{Whisker} over all the metrics. This demonstrates \app's outstanding assessment capabilities handling complex real-world teaching projects that involve visual effects and human-computer interactions. The only imperfect performance on the assignment ``Magic Show'' is due to a false positive on a single submission where the student's depicted hat sprite displays visual effects inconsistent with the physical world while the instructors think it's only a minor detail for that assignment.

We find that \textsc{Whisker} overall exhibits very low precision, which implies that it is too harsh when grading the submissions such that many correct submissions are misclassified as incorrect. In contrast, \app significantly reduced false positives. This improvement is particularly important in educational settings, where false positives undermine grading fairness and instructional trust~\cite{hsuimpautograder21}.
Additionally, \textsc{Whisker} exhibits very limited support for \texttt{ask} blocks questions that involve user interaction. Of the 30 submissions requiring user interaction, \textsc{Whisker} correctly triggers the intended conditional logic in only two cases.

\begin{table}[H]
\centering
\small
\begin{tabular}{lcccccccc}
\toprule
\textbf{Assignment} &
\multicolumn{4}{c}{\textbf{\app}} &
\multicolumn{4}{c}{\textbf{Whisker}} \\
\cmidrule(lr){2-5} \cmidrule(lr){6-9}
 & Acc(\%) & Prec & Rec & F1 & Acc(\%) & Prec & Rec & F1 \\
\midrule
Underwater World & 100 & 1.00 & 1.00 & 1.00 & 60 & 0.00 & 0.00 & 0.00 \\
Colorful Fireworks & 100 & 1.00 & 1.00 & 1.00 & 60 & 0.56 & 1.00 & 0.71 \\
Magic Show & 90.9 & 0.86 & 1.00 & 0.92 & 63.6 & 0.60 & 1.00 & 0.75 \\
Draw Five-Pointed Star & 100 & 1.00 & 1.00 & 1.00 & 50 & 0.67 & 0.57 & 0.62 \\
Ball Goes Up & 100 & 1.00 & 1.00 & 1.00 & 25 & 0.18 & 1.00 & 0.31 \\
Big Fish Eats Small Fish & 100 & 1.00 & 1.00 & 1.00 & 16.7 & 0.12 & 1.00 & 0.21 \\
Pea Shooter (Pests) & 100 & 1.00 & 1.00 & 1.00 & 16.7 & 0.09 & 1.00 & 0.17 \\
Thinking Graphics & 100 & 1.00 & 1.00 & 1.00 & 100 & 1.00 & 1.00 & 1.00 \\
Catch the Eggs & 100 & 1.00 & 1.00 & 1.00 & 100 & 1.00 & 1.00 & 1.00 \\
Hit the Bats & 100 & 1.00 & 1.00 & 1.00 & 44.4 & 0.38 & 1.00 & 0.55 \\
Mental Math Race & 100 & 1.00 & 1.00 & 1.00 & 76.9 & 0.75 & 1.00 & 0.86 \\
Math Pea Shooter & 100 & 1.00 & 1.00 & 1.00 & 81.8 & 0.86 & 0.86 & 0.86 \\
Rock--Paper--Scissors & 100 & 1.00 & 1.00 & 1.00 & 84.6 & 0.83 & 1.00 & 0.91 \\
\midrule
\textbf{Average} & \textbf{99.3} & \textbf{0.99} & \textbf{1.00} & \textbf{0.99} &
\textbf{56.8} & \textbf{0.53} & \textbf{0.89} & \textbf{0.66} \\
\bottomrule
\end{tabular}
\caption{Binary correctness comparison between \app and \textsc{Whisker} on a curated dataset consisting of submissions collected from real Scratch classes.}
\label{tab:binary-results}
\end{table}

\begin{figure}
    \centering
    \includegraphics[width=1\linewidth]{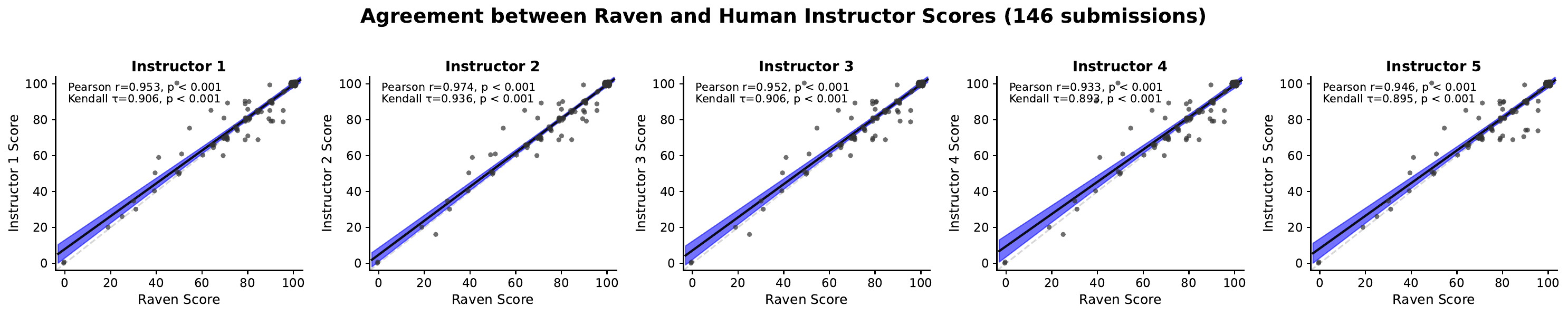}
    \caption{Agreement between \app and human instructor scores as shown in scatterplots with 146 submissions. A jittering effect with a magnitude of 1.12\% of the data range is applied on overlapping data points for visualization. Each scatterplot (\app against each instructor) exhibits strong correlations with fitted line close to y=x.}
    \label{fig:grading}
\end{figure}

\para{Raw Scores Grading Results: \app vs. Human Instructors.}
Compared to \textsc{Whisker}, \app additionally enables the provision of detailed scores based on custom grading rules. To assess the agreement in the raw scores provided by \app and human instructors, we compute the Pearson correlation (linear relationship) and Kendall's tau (non-parametric) between \app and each one of the five human instructors over the same submission dataset, as shown in the scatterplots in Fig.~\ref{fig:grading}. We see that the scores provided by \app strongly positively correlate with the scores given by the instructors, and are well calibrated. While there can be minor grading inconsistencies between different instructors, these comparisons show that \app overall aligns very well with the grading criteria of experienced Scratch instructors.

\begin{tcolorbox}[
  colback=gray!10,
  colframe=gray!50,
  boxrule=0.5pt,
  arc=2pt,
  left=6pt,
  right=6pt,
  top=6pt,
  bottom=6pt
]
\textbf{Summary for RQ1.}
In an evaluation dataset constructed from real Scratch classes, \app achieves very high grading accuracy, significantly outperforming existing the state-of-the-art Scratch assessment tool. The raw graded scores it provides closely align with experienced human instructors. Unlike prior tools that only support binary grading results, \app provides detailed grading evidence and score breakdowns, addressing a key limitation of existing Scratch assessment tools.
\end{tcolorbox}

\subsection{RQ2: Classroom Deployability}

\para{Classroom Study Design.}
To assess \app's real-world deployability, we conducted a classroom study led by 10 experienced Scratch instructors at the same after-school education center. The study spanned five class sessions, with in total 30 student participants aged 8--12 with varying Scratch experience (0--3 years).

The instructors selected and configured two assignments aligned with their teaching goals (``Ball Goes Up'' and ``Big Fish Eats Small Fish''). The instructors illustrate the end-to-end process, including assignment description, goal of the assignments, use of tool, grading criteria, as well as clarifications of questions on the survey. Students submitted their projects via USB drives, and all the grading was performed live during class. After receiving the grading results, students completed the survey. The instructors completed their survey after all the sessions were finished.

When  designing the questionnaires, we adopt the Technology Acceptance Model (TAM)~\cite{TAM89} that evaluates perceived usefulness, perceived ease of use, and willingness of adoption towards \app. TAM is a widely accepted framework for modeling user adoption of information technologies and has been widely applied in the evaluation of educational systems~\cite{deiner2024nuzzlebug, ALFRAIHAT202067, chan_students_2023, SCHERER201913, andrinaTAMslr19}. TAM characterizes tool acceptance through two major axes. Perceived usefulness (PU) measures the extent to which users believe that a system improves their task performance (e.g., ``\emph{\app helps me evaluate student Scratch projects more efficiently},'' instructor questionnaire Q3). Perceived ease of use (PEOU) captures the extent to which users perceive the system as requiring minimal effort (e.g., ``\emph{The entire process of submission and viewing results is smooth, with no technical obstacles},'' student questionnaire Q8). We also included survey questions measuring the Intention to Use and Output Quality, similar to the ones in ~\cite{deiner2024nuzzlebug}. Additionally, we added open feedback questions at the end of the questionnaires in case there are critical missing points. The full list of questions and responses from the surveys can be found in the dataset we release.

\begin{figure}
    \centering
    \includegraphics[width=1\linewidth]{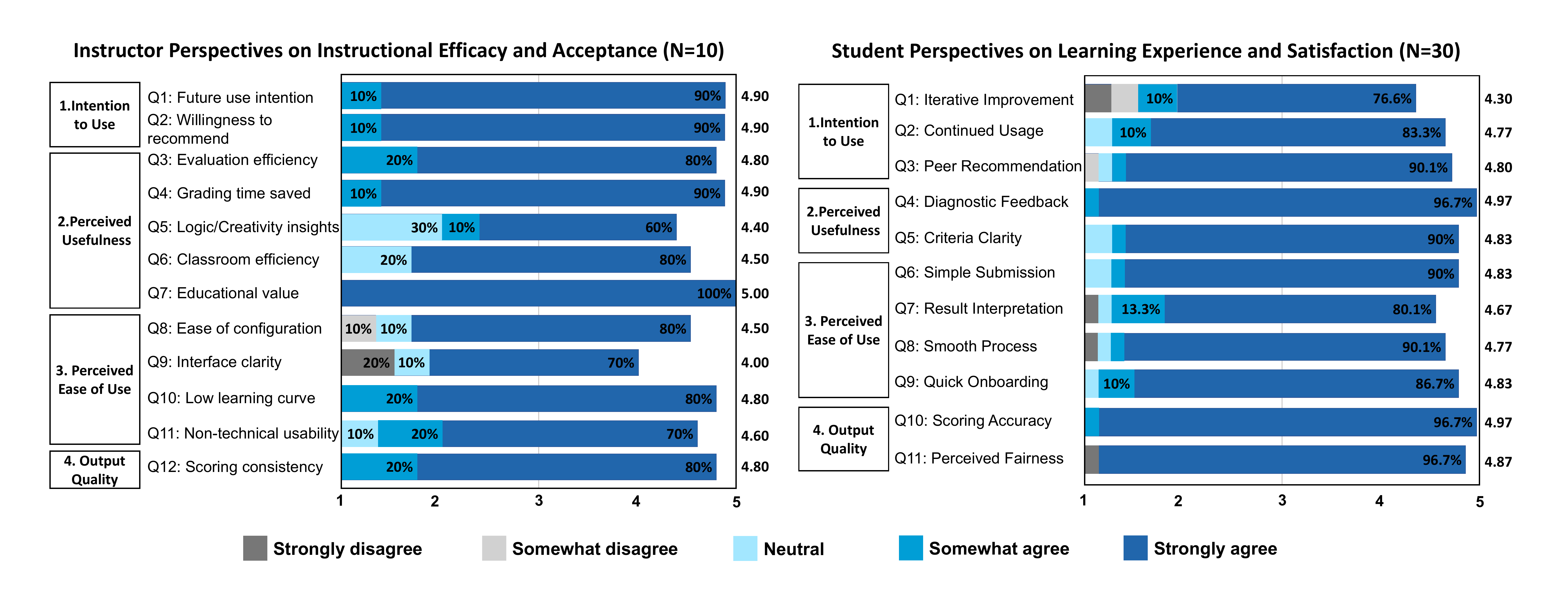}
    \caption{Survey results collected after a live classroom study conducted at an after-school education center. $N$ denotes the number of participants. The diverging stacked bar charts illustrate instructors' (left) and students' (right) feedback collected by questionnaires designed according to the Technology Acceptance Model (TAM). The questions cover Intention to Use, Perceived Usefulness, Perceived Ease of Use, and Output quality. Users respond on a Likert scale of 1-5, with the average score for each question shown on the right. The full list of questions and responses can be found in the dataset that we release.}
    \label{fig:survey-results}
\end{figure}

\para{Instructor Feedback.}
\app receives high average survey scores across all the dimensions, as shown on the left of Fig.~\ref{fig:survey-results}. Notably, they all agree that \app brings educational value and are willing to use the tool in the future. All the instructors agree that the system improves grading efficiency, and five instructors highlight in open feedback questions that \app is particularly helpful in large classrooms over 20 students, justifying the potential of scalability in real-world teaching environments. All the instructors also affirm the consistency and fairness of the grading results. In open feedback questions eight instructors express strong intent to adopt the system once available. Three instructors recommend further simplifying the task configuration and providing a web-based deployment, which we consider as an important direction for future work to improve user experience.

\para{Student Feedback.} The student survey results after using \app are shown on the right of Fig.~\ref{fig:survey-results}, which again show overall high ratings across all the dimensions. \app receives a high average score of 4.90/5 in the Perceived Usefulness dimension. Specifically, all the students agree that the grading results are reliable and that the detailed feedback help them understand the strengths and weaknesses of their submissions. It also receives a high average score of 4.78/5 in the Perceived Ease of Use dimension.
96.7\% of the students suggest that \app is a fairer system compared to manual grading, and 93.3\% of them indicate willingness to recommend it to their peers. These responses suggest strong acceptance and practical viability in real classrooms. We analyze the few 1-point ratings to identify potential weaknesses. Based on the suggestions in the open feedback questions on the student survey, the dissatisfaction is primarily due to the fact that currently submissions can only be graded by \app on teacher's computer which results in longer waiting time, and that the feedback generated is usually text-heavy. To address these user experience concerns from the product perspective, future versions will transition to a more intuitive, visual-centric reporting system and a web-based platform for convenient grading.

\begin{tcolorbox}[
  colback=gray!10,
  colframe=gray!50,
  boxrule=0.5pt,
  arc=2pt,
  left=6pt,
  right=6pt,
  top=6pt,
  bottom=6pt
]
\textbf{Summary for RQ2.}
Classroom survey results collected from both instructors and students indicate that \app overall receives strong positive feedback in live Scratch instruction environments, showing deployment potential in real-world Scratch programming education.
\end{tcolorbox}

\section{Discussion}
\label{sec:discuss}

\subsection{Sources of Discrepancy Between \app and Human Graders}

Although \app achieves high correlation with instructor grading, a small gap remains relative to inter-instructor agreement. We attribute this gap to two main factors.

\para{LLM Variability.}
Both logic- and video-based evaluation in \app rely on large language models and thus exhibit inherent variability. Logic-based evaluation may occasionally miss subtle errors due to overreliance on surface code patterns~\cite{keuning2019automatedfeedback}, while video-based evaluation can hallucinate visual events not present in execution~\cite{wang2024videohallucerevaluatingintrinsicextrinsic}. Such effects lead to sporadic misjudgments.

\para{Grading Philosophy in Creative Tasks.}
Discrepancies also arise from differing grading criteria for creative assignments. For instance, in the ``Magic Show'' task, \app enforces precise visual-state rules, whereas instructors adopt a more outcome-oriented view, prioritizing correct interaction behavior over exact visual fidelity. This philosophical difference naturally limits alignment with automated grading.

\subsection{Comparison with \textsc{ViScratch}}

\textsc{ViScratch}~\cite{viscratch} is a feedback generation system that catches Scratch bugs by aligning runtime videos with logic bugs in code. Although both \app and \textsc{ViScratch} analyze Scratch programs using video-based signals, they address different problems and serve distinct goals.

\para{Scope and Objectives.}
\textsc{ViScratch} assumes programs are \emph{buggy} and focuses on diagnosing faults in incorrect submissions. \app makes no such assumption: submissions may be correct or incorrect, and the system evaluates correctness for grading. As a result, \textsc{ViScratch} supports debugging scenarios, whereas \app is designed for formative and summative assessment.

\para{Execution and Outputs.}
\app actively executes programs using instructor-defined interaction sequences and evaluates runtime behavior through generated videos, enabling fine-grained, rubric-based scoring with partial credit. In contrast, \textsc{ViScratch} does not rely on systematic execution-driven video generation and primarily produces binary or categorical bug diagnoses.

\subsection{Design Implications and Future Work} 

The core value of \app lies in shifting the focus of automated assessment from \emph{how to design system-level assertions} to \emph{how to evaluate observed behavioral outcomes}. This shift enables automated grading to better serve educational objectives rather than rigid technical specifications. By simplifying and customizing the instructor-facing task configuration process through video generation rules, and by providing detailed grading evidence, \app substantially lowers the technical barrier for educators. In addition, the lightweight video understanding module effectively supports human--computer interaction scenarios, addressing the long-standing challenge posed by \texttt{ask} blocks in Scratch programs.

Future work will focus on three directions. First, we plan to develop more efficient video generation and analysis techniques to further reduce assessment latency. Second, we aim to extend the system to better support the evaluation of creative and open-ended projects that go beyond fixed grading rubrics. Third, we plan to build a cloud-based assessment platform that provides a large collection of reusable tasks for instructors and practice opportunities for students, supporting online task customization and submission, and fostering a scalable educational assessment ecosystem.

We believe that \app not only provides a practical solution for Scratch education, but also opens new avenues for automated assessment in other graphical programming environments, ultimately contributing to the quality and accessibility of computational thinking education.

\section{Threats to Validity}
\label{sec:threats}

We discuss potential threats to the validity of our evaluation and how we mitigate them.

\para{Dataset Size and Task Complexity.}
A potential threat to external validity is whether our dataset adequately represents the diversity and complexity of Scratch programs encountered in real classrooms. To address this concern, our evaluation dataset consists of 13 assignments collected from real-world instructional settings, spanning multiple grade levels, difficulty levels, and programming constructs.
The assignments provide comprehensive coverage of fundamental Scratch categories, including Motion, Looks (e.g., costumes and special effects), Sound, Events (e.g.,broadcasting), Control (e.g., loops and cloning), Sensing (e.g., timer, user interaction, and collisions), Operators (e.g., variables, randomness, and logic), and specialized extensions such as the Pen module. This diversity reduces the risk that our results are biased toward a narrow class of programs. Nevertheless, we acknowledge that the dataset size is limited, and future work should evaluate \app on larger-scale datasets.

\para{Model Stability and Nondeterminism.}
Another threat to internal validity stems from the nondeterministic behavior of large language models used in both logic-based and video-based evaluation.
We mitigate this threat in two ways. First, the video checker runs each evaluation three times and considers a grading rule as satisfied only if all the runs produce the same result, reducing the impact of sporadic hallucinations. Second, \app employs a lightweight \emph{check, repair, and merge} procedure to reconcile logic- and video-based results: when conflicts occur, video judgments take precedence, reflecting the primacy of observable runtime behavior in Scratch assessment and improving robustness to isolated model errors. 

\para{Human Grading Variability.}
There are inevitably variabilities of grading outcomes by different human instructors even on the same student submission, as each grader can have their own interpretation over the same grading rules. Our evaluation assumes grading rules accurately reflect instructional intent; poorly specified rules could limit grading validity. In our studies, we recruited multiple experienced instructors to construct the evaluation dataset. The ground truth labels in the evaluation dataset are based on a majority vote of these instructors, which ensures the quality of reference. We showed in RQ1 that \app provides grading results highly consistent with multiple experienced Scratch teachers. Also, the introduction of \app may help reduce human grading variabilities in teaching, as illustrated by the high scores \app received on scoring accuracy and perceived fairness in the student survey in RQ2.

\section{Related Work}
\label{sec:related}

\para{Automated Assessment for Scratch Programs.}
Automated assessment for Scratch programs has attracted increasing attention as block-based programming has become widely adopted in Computer Science education. The assessment of Scratch programs has evolved from static code analysis to assertion-based testing. Early approaches, such as \emph{Hairball}~\cite{boe2013hairball} and \emph{Dr. Scratch}~\cite{moreno2015dr}, pioneered \emph{static analysis} by parsing the Abstract Syntax Tree (AST) to detect code smells, bad practices, or missing blocks~\cite{aivaloglou2016code, techapalokul2017code}. While effective for assessing coding style and computational thinking concepts~\cite{romangonzalez2017ct, TANG2020103798, autoassesment22}, static methods fail to capture runtime logic and functional correctness, often leading to false positives in semantically complex projects~\cite{fppattern17}.

To address this, assertion-based testing~\cite{4222570} techniques were introduced to evaluate functional behavior. For example, \textsc{Whisker}~\cite{stahlbauer2019whisker} models Scratch programs as event-driven systems and evaluates correctness by simulating user interactions (e.g., key presses and mouse events) and checking state assertions. Subsequent work has focused on improving test generation for \textsc{Whisker}, including search-based testing~\cite{DBLP:journals/corr/abs-2009-04115}, model-based testing~\cite{götz2022modelbasedtestingscratchprograms}, and automated oracle generation~\cite{deiner2023automated}. While these techniques reduce the burden of manual test authoring, they fundamentally inherit the limitations of assertion-based evaluation, including sensitivity to execution order, reliance on precise state specifications, and difficulty handling visual outcomes and diverse implementations.

Other program analysis tools target specific tasks such as bug detection, program repair~\cite{deiner2024nuzzlebug,schweikl2025repurr,STRIJBOL2024101617,Pydex,Gmerge,EcoScratch}, or feedback generation~\cite{price2017isnap,fein2022catnip,VeriCI,rivers2017itap,Clef,NETTLE,gu2026context}, often assuming the presence of a reference implementation or known buggy submissions. These systems typically focus on \emph{debugging} rather than grading and do not address the challenge of assessing open-ended, visually defined correctness in real classroom settings.

\para{Visual and Behavior-Based Program Assessment.}
Since Scratch projects are inherently visual and interactive, their assessment shares challenges with Graphical User Interface (GUI) testing and automated game validation. 
Traditional GUI testing tools, such as \emph{Sikuli}~\cite{yeh2009sikuli} and \emph{GUITAR}~\cite{nguyen_guitar_2014}, utilize screenshot matching and widget-tree analysis to verify interface correctness~\cite{chang2010gui, BANERJEE20131679,droidbot17}. In the domain of game development, automated agents have been employed to explore game states~\cite{zheng2019wuji, alhedat2020automated} or detect glitches via visual artifacts~\cite{ag3liang23}. However, most existing techniques still depend on predefined assertions or task-specific heuristics and are not designed for open-ended educational programs with high implementation diversity.
Recent work has begun to explore video as an analysis artifact for understanding program execution~\cite{ScratchEval}, particularly in robotics and embodied agents~\cite{pmlr-v80-sun18a,NEURIPS2021_f8b7aa3a}. These efforts demonstrate that video captures aspects of system behavior that are difficult to express through program states alone.
Nevertheless, existing visual analysis tools are primarily designed for \emph{crash detection} or \emph{debugging}~\cite{viscratch,electronics12112382,pham_tailoring_2014,cv2se_survey22} rather than pedagogical grading. they do not address instructor-facing configuration or real-world educational deployment.

\para{Large Language Models in Educational Assessment.}
Large language models (LLMs) have recently been applied to educational tasks such as automated feedback generation~\cite{chen2021codex,phung2023feedback}, short-answer grading~\cite{Chang_Ginter_2024,grevisse_llm-based_2024}, and programming assistance~\cite{codeaid24}. In the context of programming education, LLMs have been used to explain code~\cite{llm_understand_code24,llmvsstudent_codeexpl23}, identify bugs~\cite{teachableagentbug24}, or guide students through debugging steps~\cite{Stitch}. These approaches primarily operate on source code or textual artifacts and assume well-defined specifications or reference solutions.

More recently, multimodal LLMs like \texttt{GPT-4V}~\cite{openai2024gpt4technicalreport} and \texttt{Qwen-VL}~\cite{bai2025qwen3vltechnicalreport} have shown promise in reasoning over visual inputs, including diagrams and videos~\cite{yang2023dawnlmmspreliminaryexplorations,wang2024qwen2vlenhancingvisionlanguagemodels}. However, their application to automated assessment of graphical, event-driven programs remains largely unexplored. In particular, existing work does not leverage LLMs to reason over execution videos as a first-class grading artifact, nor does it integrate video-based reasoning with instructor-defined grading rules.

\section{Conclusion}
\label{sec:conclusion}

This work introduces \app, a significant advance in automated assessment for Scratch programs. By integrating large language models with video-based analysis, \app addresses key limitations of prior approaches, enabling evaluation of visual behaviors, accommodation of diverse student implementations, and support for human–computer interaction assessment. Our evaluation shows that \app produces accurate, fine-grained assessments that align closely with instructor grading, and in-classroom deployments show that \app is practical and usable in real-world settings. These results suggest that \app provides a strong foundation for scalable and grounded assessment in block-based programming environments.


\bibliographystyle{ACM-Reference-Format}
\bibliography{scratch}

\end{document}